\documentclass[runningheads]{llncs}
\usepackage[american]{babel}
\usepackage[T1]{fontenc}
\usepackage{times}
\usepackage{caption}
\usepackage[]{algorithm2e}
\usepackage{textcomp}
\usepackage{epsfig,graphicx}
\usepackage{xcolor}
\usepackage{amsfonts,amsmath,amssymb}

\usepackage{booktabs}
\usepackage{longtable}
\usepackage[utf8]{inputenc}
\usepackage{supertabular}
\usepackage{listings}
\usepackage{cite}
\usepackage{url}
\usepackage{float}
\definecolor{backcolour}{rgb}{0.95,0.95,0.92}
\begin{document}

\title{WebChecker: A Versatile EVL Plugin for Validating HTML Pages with Bootstrap Frameworks}
\author{Milind Cherukuri}
\authorrunning{M.C}
\institute{\email{cherukurimilind@gmail.com}}

\maketitle

\begin{abstract}
WebChecker is a plugin for Epsilon Validation Language (EVL), designed to validate both static and dynamic HTML pages utilizing frameworks like Bootstrap. By employing configurable EVL constraints, WebChecker enforces implicit rules governing HTML and CSS frameworks. The effectiveness of the plugin is demonstrated through its application on Bootstrap, the widely adopted HTML, CSS, and JavaScript framework. WebChecker comes with a set of EVL constraints to assess Bootstrap-based web pages. To substantiate the claims, I present an illustrative example featuring two solutions that effectively enforce implicit rules. 
\end{abstract}

\section{Introduction}
In the current state of web development, many HTML and CSS frameworks come with implicit rules. Such frameworks include Bootstrap, Materialize \cite{materialize}, Foundation \cite{foundation}, and uikit \cite{uikit}. These rules are developed for better use of such frameworks and best practices. Therefore, not following such rules impacts appearances and framework features. Most of these frameworks are built to be responsive to different screen sizes, such as large screens of personal computers and small screens of smart devices. To provide a consistent look and feel on all devices, it is important to follow such implicit rules.

The Accessible Rich Internet Applications (ARIA) \cite{aria} specification introduces a handful number of rules for web pages in order to make them accessible for users with disabilities through assistive technologies such as screen readers. Such assistive technologies require web pages to have certain attributes defined or enabled on HTML elements. Bootstrap, for example, provides default classes and ARIA attribute values throughout the framework for such technologies. Therefore, conforming to such rules makes web pages available to a greater audience.

However, the caveat is to read the frameworks' documentation carefully to capture such constraints and rules. This could be quite tedious and expensive work. WebChecker is built to capture such rules so that users of this framework can easily check their web pages for conformance with such regulations. While WebChecker is built with HTML frameworks in mind, it could be easily used for web pages built without such frameworks. However, in this paper, I focus on the Bootstrap framework. Furthermore, WebChecker is able to check static and dynamic pages, which are generated by scripting languages such as PHP and JavaScript. 

The rest of the paper is structured as follows: Section II explains the problem, the current state-of-the-art solution, and the limitations that motivated this project. In Section III, I explain the WebChecker plugin and show our solution to the problem. In Section IV, I present a few sample implicit rules captured for the Bootstrap framework. I compare the current solution and our solution of the problem and its improvements in Section V. In Section VI, I discuss related work. Lastly, Section VII concludes the paper and outlines the future of the plugin and future work to enhance the plugin.  

\section{Background and Motivation}

Frameworks such as Bootstrap come with a handful number of implicit rules. While developing this project, I captured an initial 25 rules for the Bootstrap framework. These rules should be followed in order to use the framework properly. These rules are created by the developers of such frameworks and explained in the framework’s documentation in a natural language. Users of such frameworks should carefully read the documentation in order to understand how to use the framework. This process can be quite long and tedious, expensive, and prone to errors. 

\begin{lstlisting}[language=HTML, caption=Bootstrap Grid Example, label={lst:bge}]
<!-- sample.html -->
<div class="container">
  <div class="row">
    <div class="col">
      Column One
    </div>
    <div class="col">
      Column Two
    </div>
    <div class="col">
      Column Three
    </div>
  </div>
</div>
\end{lstlisting}

Given a set of rules, currently, there is not a straightforward process to check if they are enforced against the HTML pages. At least, there is no easy way that requires minimum effort and very little code. After collecting the rules, the current solution flow is as follows:
1. Read and encode the HTML page. 
2. Translate each rule into a method. 
3. Check the rule against the HTML page.
4. Repeat 1-3 for every rule.
For example, in Bootstrap, any content should be under a \textit{<div>} element with class \textit{col}, which should be under a \textit{<div>} element with class \textit{row}, which should be under another \textit{<div>} element with class \textit{container}. The HTML code sample is at Listing \ref{lst:bge}.

To enforce this rule, the current state-of-the-art solution is given in Listing \ref{lst:ebgr}. Since the plugin is written in Java, I chose it as our programming language of choice to implement this solution. However, any programming language can be used. Some of these might require more or less coding. Listing \ref{lst:ebgr} shows that for enforcing the rule in Listing \ref{lst:bge}, the code can become lengthy and unreadable, which makes the logic confusing to reason about. It is also difficult to reuse and maintain this code since each rule has its own requirements and structure. Depending on the rules, it is likely to have more nested \textit{if statements}.
Furthermore, the solution is not configurable. That is, it is difficult to make changes without breaking the code logic. This approach requires framework developers and users to spend quite some time writing similar code as Listing \ref{lst:ebgr} to enforce rules and constraints and write documentation for such code.   

The above difficulties have motivated us to implement a plugin, WebChecker, for the Epsilon Validation Language (EVL) \cite{Kolovos2009}.  This plugin is designed to enforce rules and constraints for any framework, such as Bootstrap. With WebChecker, 1) a static or dynamic HTML file, 2) an HTML web page through its URL, and 3) a specific section of an HTML page can be checked for conformance. This plugin provides EVL constraint reusability across multiple projects and improves readability. Framework users and developers are able to focus on the rules and constraints. The next section explains WebChecker in more detail.    

\begin{lstlisting}[language=Java, caption=Enforcing Bootstrap Grid Rule, label={lst:ebgr}]
		File input = new File("files/bootstrap/newCheck.html");
		try {
			Document doc = Jsoup.parse(input, "UTF-8");
			Elements elements = doc.getElementsByTag("div");
			for (Element element : elements) {
				if (element.hasClass("col-sm-4")) {					
					if (!element.parent().hasClass("row")) {
						System.out.println("A div element with class col should have a parent element with class row");
					} else {
						if (!element.parent().parent().hasClass("container")) {
							System.out.println("A div element with class col should have a parent element with class row, which has a parent with class container.");
						}
					}
				}
			}		
		} catch (IOException e) {
			e.printStackTrace();
		}
\end{lstlisting}

\section{WebChecker Plugin}
\begin{figure}
  \includegraphics[width=\linewidth]{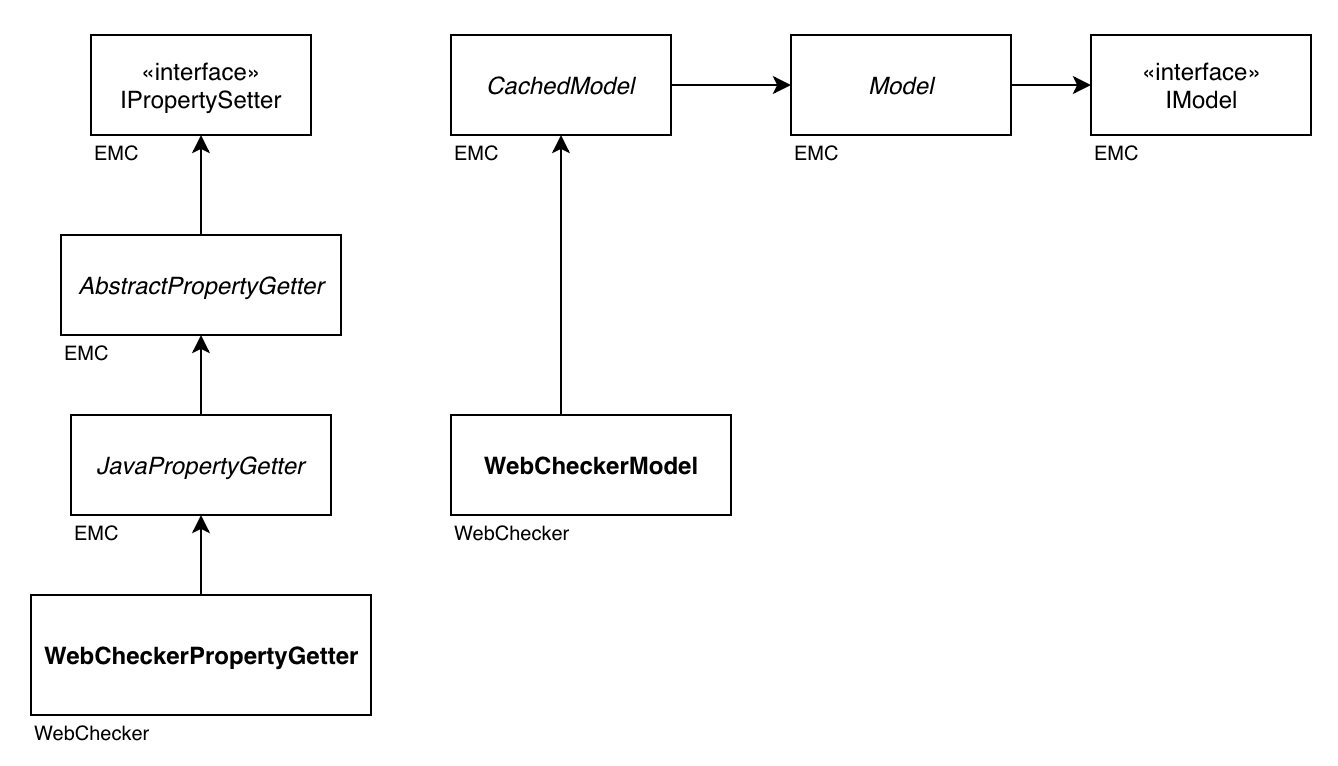}
  \caption{WebChecker and Partial EMC Diagram}
  \label{fig:webcheckerclass}
\end{figure}

Epsilon Validation Language (EVL) \cite{Kolovos2009} is one of the languages of Epsilon \cite{EpsilonProject}, which is an Eclipse project that provides languages for model management such as model validation, model transformation, code generation, pattern matching, model merging, etc. In the context of WebChecker, a model is an HTML page or a section of the page. WebChecker implements an Epsilon Model Connectivity (EMC) layer that provides an interface, IModel, to capture the HTML model and EVL constraint file. The two main classes implemented by the plugin are \textit{WebCheckerModel} and \textit{WebCheckerPropertyGetter} While Epsilon provides the other classes. Figure \ref{fig:webcheckerclass} shows this relation. Furthermore, EVL users can use the features of WebChecker without setting up a new development workflow or installing new software.  

In the following subsections, I present our solution and explain the WebChecker EVL structure.

\subsection{WebChecker}
The WebChecker flow, after collecting the constraints, is: 
1) Write an EVL constraint.
2) Choose the HTML source (i.e., an HTML file, URL, or section of HTML).
These two steps are modular and readable. Listing \ref{lst:webcheckerSolution} is an example of an EVL file, which shows the solution for Listing \ref{lst:bge}. 

WebChecker follows the \textit{separation of concern}\footnote{According to Wikipedia \textit{separation of concern} is as a design principle for separating a computer program into distinct sections.} principle with writing minimum code. In particular, there is an EVL source file and an HTML source, which could be an HTML file content or its URL. The EVL file that contains the constraints such as Listing \ref{lst:webcheckerSolution} is run against the HTML source. Listing \ref{lst:linkEVLAndHTML} shows this connection. Lines 10 and 11 of Listing \ref{lst:linkEVLAndHTML} set the HTML and EVL source, respectively. Moreover, line 15 of Listing \ref{lst:linkEVLAndHTML} shows how to handle errors returned from the EVL constraints. This way, the framework users and developers only need to focus on constraints that are not satisfied.   

\begin{minipage}{\linewidth}
\begin{lstlisting}[language=Java, caption=Checking an EVL file against an HTML source file, label={lst:linkEVLAndHTML}]
 	//sample.java    
		//Step 1: Get the source to be validated
		String html = "files/bootstrap/newCheck.html";
		
		//Step 2: Write your validation using Epsilon Validation Language
		String evl = "files/bootstrap/newCheck.evl";
		
		//Step 3: Check the validation against the html in step 1
		WebChecker checker = new WebChecker();
		checker.setSource(html);
		checker.setValidation(evl);
		checker.check();
		
		//Step 4: Check the result
		List<String> errors = checker.errors();


\end{lstlisting}
\end{minipage}

\subsection{WebChecker EVL Structure}
The WebChecker EVL structure follows the same EVL structure as in \cite{Kolovos2009}. However, I will only explain the current syntax used by WebChecker.

\begin{lstlisting}[language=Java, caption=Enforcing Bootstrap Grid Rule by Using WebChecker, label={lst:webcheckerSolution}]
//sample.evl
context t_div { 
    constraint DivWithColHasRowParent {
        guard : self.class.includes("col-*")
        check : self.parent.hasClass("row") and self.parent.is("div")
        message : "A <div> element with class col should have a parent <div> element with class row."
    }
    constraint DivWithRowHasContainerParent {
        guard : self.class.includes("row")
        check : self.parent.hasClass("container") and self.parent.is("div")
        message : "A <div> element with class col should have a parent <div> element with class row."
    }
}

\end{lstlisting}

\begin{itemize}
\item \textit{Context}: an EVL file could have one or more contexts. To capture an HTML element, a context should start with \textit{t\_}, for \textit{type}, followed by the name of the HTML element such as \textit{t\_div}, \textit{t\_picture}, \textit{t\_section}, \textit{t\_button}, etc.
\item \textit{Constraint}: each \textit{Context} could have one or more constraints. A \textit{constraint} has a name that is used to differentiate each constraint from each other. While the constraint name does not matter, I suggest having a name that identifies the rule. Each \textit{constraint} has three blocks \textit{guard}, \textit{check}, and \textit{message}. Optionally, an EVL constraint could have a \textit{fix} section, where the model is fixed if the constraint is not satisfied. Currently, WebChecker does not utilize this feature. Section VII explains future enhancements to WebChecker where the \textit{fix} block is explained. 
\item \textit{guard}: this block captures a specific section of the context for checking. \textit{guard} must return a boolean value. If true, the \textit{check} block is executed. The \textit{guard} block accepts compound boolean statements by using \textit{and} and \textit{or}.      
\end{itemize}
\begin{itemize}
\item \textit{check}: after the \textit{guard} block is satisfied, this block is executed to \textit{check} if the constraint is satisfied. Expressions in this block must return a boolean value. If the expression returns true, the constraint is satisfied, and hence the \textit{message} block is not executed. Similarly to the \textit{guard} block, this block accepts compound boolean statements.
\item \textit{message}: this block returns a message if the constraint is not satisfied. WebChecker provides a convenient method to capture all unsatisfied constraints' messages shown in Line 15 of Listing \ref{lst:webcheckerSolution}.  
\end{itemize}

\section{Bootstrap Implicit Rules by WebChecker}
WebChecker enforces an initial 25 rules for the Bootstrap framework. There are much more rules to be captured. Here, I only show and explain five randomly chosen implicit rules for the framework. For the complete list of laws enforced by the plugin please see the project's GitHub repository\footnote{https://github.com/tebinraouf/webchecker}. For brevity, I do not show the actual EVL syntax.

\begin{itemize}
\item \textit{ScreenReaderButton}: this constraint captures buttons with class \textit{close} and validates if the \textit{<button>} element has the \textit{aria-label} attribute defined for assistive technologies.
\item \textit{AlertLinkInDivAlert}: this constraint captures \textit{<a>} elements with class \textit{alert-link} and validates if the parent element includes \textit{alert} and \textit{alert-*} classes. The asterisk (*) is a wild card such as \textit{alert-success} or \textit{alert-danger} etc.
\item \textit{BtnGroupToggle}: this constraint captures \textit{<div>} elements with class \textit{btn-group-toggle} and checks if its \textit{data-toggle} attribute is defined and its value is equal to \textit{buttons}.
\item \textit{BadgeClassSiblingRelation}: this constraint captures \textit{<span>} elements with \textit{badge} and \textit{badge-*} classes and checks if either the previous sibling element or the next sibling element has a \textit{sr-only} class. This is required for screen readers to know what the badge represents rather than what it looks like.
\item \textit{ImageInPictureWithImgClass}: this constraint captures  \textit{<img>} elements whose parent is the  \textit{<picture>} element and checks if the \textit{<img>} element has a \textit{img-*} class. This is required since a <picture> element indicates that there is a picture in the tag.
\end{itemize}

\section{Improvements over Current Solution}
As shown from the example above, WebChecker is built to be easy, modular, readable, and configurable. These features address and solve most of the challenges developers face while using a program analyzer as studied by Microsoft researchers \cite{christakis16}. WebChecker, specifically, offers custom warning messages, platform independence, static and dynamic page analysis, and easy integration into the development workflow. Importantly, it is left to the developer to check HTML pages against any set of EVL constraints, which makes WebChecker stand out among other validators that have default un-configurable rules.    

Furthermore, WebChecker provides an abstract class, \textit{WebChecker},  to interact with the underlying model. While users can (should) use the \textit{WebChecker} class, they are not restricted to using the \textit{WebCheckerModel} class, which implements the CachedModel of EMC. While \textit{WebCheckerModel} requires background knowledge of Epsilon, it could be very powerful and configurable. The WebChecker abstract class allows a user to focus on the constraints rather than writing code, which saves time and resources. WebChecker's modularity will enable it to be reusable and extensible. If new framework rules are in place, WebChecker can be modified or extended to support such regulations. 

Code written like Listing \ref{lst:ebgr} is prone to errors and confusion, which makes the code unreadable to non-technical stakeholders such as managers and web designers. Also, as stated above, the code in Listing  \ref{lst:ebgr}  is written in Java. It is likely the developer uses a language that their workflow uses for validation, such as JavaScript, PHP, C\#, Python etc. This makes it difficult to use the code in another development environment that uses different programming languages and technology stacks. In contrast, WebChecker's EVL constraints are easy to reason about and independent of the development environment. That is, as long as the source HTML is given to WebChecker, how the HTML is generated does not matter. 

Line 3 of Listing \ref{lst:ebgr} uses \textit{jsoup: Java HTML Parser}\cite{jsoup} library to parse the HTML file into \textit{jsoup's Element} objects. While \textit{jsoup} is very powerful, it is irrelevant in this context because a user does not need to learn a new library that is unrelated to a validation task. How the HTML file is parsed into objects should be done internally. WebChecker abstracts this by having the user only set the source of the HTML file as shown in Line 10 of Listing \ref{lst:linkEVLAndHTML}.

The \textit{for-loop} and the \textit{try-catch} blocks in Listing \ref{lst:ebgr} are the least important to the user. WebChecker improves this by introducing a clear, readable structure, as explained in the WebChecker EVL Structure section above. Table \ref{table:compareWebChecker} shows a side-by-side comparison of WebChecker with other current validators, purely based on our research and experiment.    
 
\begin{table}[H]
\centering
\caption{WebChecker Plugin Compared to Other Tools}

\begin{tabular}{|l|c|c|}
\hline
\textbf{Features}   	 & \textbf{WebChecker} 	& \textbf{Other} \\ \hline
Reusability 	 & \checkmark   	&       \\ \hline
Configurable      & \checkmark        			&       \\ \hline
Extensible        &    \checkmark        			&       \\ \hline
Platform Independent        &    \checkmark        			&       \\ \hline
Minimum Coding        &    \checkmark        			&       \\ \hline	
Task Specific       &    \checkmark        			&       \\ \hline
Dynamic Page Analysis       &    \checkmark        			&       \\ \hline
Static Page Analysis       &    \checkmark        			& \checkmark      \\ \hline	
\end{tabular}
\label{table:compareWebChecker}
\end{table}

\section{Related Work}
Bootlint \cite{bootlint} is a Bootstrap-specific linter for checking common HTML mistakes in web pages built in a vanilla way; that is, pages where default Bootstrap classes are used. Bootlint uses JavaScript to check for Bootstrap rule conformance, which is not configurable. The source code of Bootlint, which is in the \textit{bootlint.js}\footnote{https://github.com/twbs/bootlint/blob/master/src/bootlint.js} file of the Bootlint project, is much lengthier than the code in Listing 2. This validates our points on the current limitations of the current solution, as explained above. 

Policheck is Microsoft's internal tool that checks codes, comments, and content, including web pages. For example, it can check if a web page contains inappropriate content or if its style matches best practices and guidelines \cite{christakis16}.

ESLint \cite{eslint} is a static JavaScript analysis linting utility written in JavaScript to enforce coding styles through pluggable rules. With ESLint, developers can analyze JavaScript code by using default rules or creating new rules without running the JavaScript code. ESLint requires a configuration file in JavaScript Object Notation (JSON), YAML (YAML Ain't Markup Language), or JavaScript. Most importantly, the configuration file includes the source of the rule file, which is a JavaScript file that has the implementation of the rule. While ESLint is powerful and popular, the JavaScript code for each rule is not reusable.

\section{Conclusion and Future Work}

In this paper, I propose a new plugin for the Epsilon Validation Language with a concrete example. I identify the current solution developers use to check web pages for framework rule conformance and compare our solution with its improvements. 

I believe that with web pages being ubiquitous, there will be more frameworks with implicit rules and less detailed documentation. If developers of these frameworks have a tool such as WebChecker, they can write exhaustive lists without providing much documentation since EVL constraints are easily readable and close to natural language. 

Further research on the topic includes extending WebChecker to support fixes when a constraint is not satisfied, evaluating other framework rules and capturing such implicit rules, providing WebChecker through popular text editors, validating web pages online, project-wise validation, and developing a domain-specific language for capturing and enforcing such regulations that is easy to integrate with a development workflow.

\end{document}